\documentclass[12pt]{article}
\usepackage{putex}
\usepackage{graphicx}

\newcommand{\BesselK}[1]{{\bf K_{#1}}}
\newcommand{\BesselI}[1]{{\bf I_{#1}}}
\newcommand{\StruveL}[1]{{\bf L_{#1}}}

\providecommand{\abs}[1]{\lvert#1\rvert}

\begin{document}

\preprint{PUPT-2236 \\ LMU-ASC 34/07}

\institution{PU}{Joseph Henry Laboratories, Princeton University, Princeton, NJ 08544, USA}
\institution{MaxPlanck}{Ludwig-Maximilians-Universit\"at, Department f\"ur Physik, Theresienstrasse 37, \cr 80333 M\"unchen, Germany}

\title{Energy disturbances due to a moving quark from gauge-string duality}

\authors{Steven S. Gubser,\worksat{\PU,}\footnote{e-mail: {\tt ssgubser@Princeton.EDU}}
Silviu S. Pufu,\worksat{\PU,}\footnote{e-mail: {\tt spufu@Princeton.EDU}} and
Amos Yarom\worksat{\MaxPlanck,}\footnote{e-mail: {\tt yarom@theorie.physik.uni-muenchen.de}}}

\abstract{Using AdS/CFT, we calculate the energy density of a quark moving through a thermal state of ${\cal N}=4$ super-Yang-Mills theory.  Relying on previous work for momentum-space representations as well as asymptotic behaviors, we Fourier transform to position space and exhibit a sonic boom at a speed larger than the speed of sound.  Nontrivial structure is found at small length scales, confirming earlier analytical work by the authors.}

\PACS{}
\date{June 2007}

\maketitle

\section{Introduction}
Heavy ion collision experiments at the Relativistic Heavy Ion Collider (RHIC) are believed to probe properties of a deconfined state of hadronic matter, the Quark Gluon Plasma (QGP) \cite{Arsene:2004fa,Adcox:2004mh,Back:2004je,Adams:2005dq}.  The phenomena of jet-quenching and jet-splitting observed at RHIC are challenging to understand theoretically.  A review of theoretical treatments of jet-quenching can be found, for example, in \cite{Majumder:2007iu}.  Theoretical works on jet-splitting include \cite{Chiu:2006pu,Armesto:2004pt,Vitev:2005yg,Polosa:2006hb,Koch:2005sx,Casalderrey-Solana:2006sq}.  Clearly it is desirable to study similar phenomena in strongly coupled $\mathcal{N}=4$ super-Yang-Mills theory (SYM), where substantial progress can be made via AdS/CFT starting from calculations in supergravity.  While direct comparisons between SYM and QCD are fraught with uncertainties,  such calculations at least provide an independent perspective on dissipative effects arising from a quark propagating through a strongly coupled thermal medium.  In this paper the moving quark is infinitely massive, pointlike, and fundamentally charged.

A heavy quark moving at constant velocity in an infinite, static, thermal bath of ${\cal N}=4$ super-Yang-Mills theory can be described in the dual gravity picture by a string whose endpoint is on the boundary of the AdS${}_5$-Schwarzschild spacetime.  This string is responsible for the drag force on the quark \cite{Herzog:2006gh,Gubser:2006bz} (see also the closely related work \cite{Casalderrey-Solana:2006rq}, as well as other work \cite{Liu:2006ug} on jet-quenching in AdS/CFT).  The configuration mentioned has a steady-state approximation built in: the trailing string is moving at the same velocity $v$ as the quark.  Our convention is to take this velocity to be in the $x_1$ direction: $\vec{v} = (v,0,0)$.  We employ mostly plus metric, so that $x^1 = x_1$ but $x^0 = -x_0 = t$.

The string perturbs the geometry of AdS${}_5$-Schwarzschild in a manner explained in \cite{Friess:2006fk}. From the asymptotics of those metric perturbations one may extract the expectation value of the gauge theory stress-energy tensor.  In \cite{Friess:2006fk} the Fourier coefficients of this stress-energy tensor have been evaluated.  It is clearly desirable to pass to a position space description: this is a necessary step before hadronization can be considered, and it also helps one's intuitions about the dissipative mechanisms involved.  The main result of this paper is the computation of the energy density at a fixed time as a function of position $\vec{x}$.  More precisely, we compute a dimensionless quantity
 \eqn{EDef}{
  {\cal E}(\vec{X}) = {\sqrt{1-v^2} \over (\pi T)^4 \sqrt{g_{YM}^2 N}}
    \left( \langle T_{00}(0,\vec{x}) \rangle -
      {3\pi^2 \over 8} N^2 T^4 \right) \,.
 }
Here $\langle T_{00}(0,\vec{x}) \rangle$ is the thermal expectation value of the total energy density in the gauge theory at time $t=0$. For convenience, an overall factor of $\sqrt{1-v^2}$ has been brought out explicitly in \eno{EDef}.  We always use spacetime coordinates such that the plasma is at rest.  The second term inside the parentheses in \eno{EDef} is the contribution to $\langle T_{00} \rangle$ from the thermal bath.  We will generally work with dimensionless position space and Fourier space variables:
 \eqn{KXDef}{
  \vec{X} = \pi T \vec{x}, \qquad\quad
   \vec{K} = \vec{k} / \pi T \,.
 }

With the Fourier coefficients ${\cal E}(\vec{K})$ in hand, the problem we face is simply to compute a three-dimensional Fourier transform:
 \eqn{FTDef}{
  {\cal E}(\vec{X}) \equiv \int {d^3 K \over (2\pi)^3}
    e^{i\vec{K} \cdot \vec{X}} {\cal E}(\vec{K}) \,.
 }
A difficulty in carrying out this Fourier transform is that ${\cal E}(\vec{K})$ grows linearly with $\vec{K}$ and is also singular at small $\vec{K}$.  Our strategy will be to find asymptotic expressions with known analytic Fourier transforms at large and small wave-numbers, subtract them away, and pass the remainder through a fast Fourier transform (FFT)\@.  The simplest of these expressions is the Coulombic near-field of the quark, which is proportional to $1/x^4$ in the rest frame of the quark and takes the form
 \eqn{ECoulombX}{
  {\cal E}_{\rm Coulomb} (\vec{X}) = {(1-v^2)^{5/2} \left[X_1^2 + (1+v^2) X_\perp^2\right]\over 12 \pi^2 \left[X_1^2 + (1-v^2) X_\perp^2\right]^3}
 }
in the rest frame of the plasma.  Here $X_\perp = \sqrt{X_2^2 + X_3^2}$.  Finite-temperature corrections to \eno{ECoulombX} which are still singular at $X=0$ were found in \cite{Yarom:2007ap,Gubser:2007nd,Yarom:2007ni} and are developed further in this work.  These corrections may be interesting in their own right, because they provide some gauge-invariant information about the energy loss mechanisms close to the quark.  They might be used to seed hydrodynamical simulations.

In presenting our results we will usually consider the quantity
\begin{align}
\label{SuperSubtracted}
  E(\vec{X}) &\equiv {\cal E}(\vec{X}) -
   {\cal E}_{\rm Coulomb}(\vec{X}) \\
\notag
   &=\left(\substack{\text{\small{total}} \\ \text{\small{energy density}} \\ \text{\small{of system}}}
    -\substack{\text{\small{energy density}} \\ \text{\small{of plasma}} \\ \text{\small{in equilibrium}}}\right)
    -\substack{\text{\small{energy density}} \\ \text{\small{of moving quark}} \\ \text{\small{in vacuum}}}\,.
\end{align}
Up to an overall prefactor (see \eno{EDef}), $E(\vec{X})$ is the total energy density of the moving quark and the thermal bath, minus the energy density of the thermal bath in the absence of the moving quark, minus the energy density of the moving quark in the absence of the thermal bath, all at a fixed time. We will refer to $E(\vec{X})$ somewhat loosely as the net energy density.  It may be positive or negative.

The reader wishing to skip over technical details can find a summary of our results in section~\ref{NUMERICAL}.  Briefly, we find good agreement both with the analytical estimates \cite{Yarom:2007ap,Gubser:2007nd,Yarom:2007ni} at small length scales, and with linearized hydrodynamics at large length scales.

During the late stages of this project, we learned of a similar study of the energy density which has some overlap with this work \cite{YaffeToAppear}.

\section{Real space calculation of the energy density}
\label{REVIEW}

In earlier work \cite{Friess:2006fk,Gubser:2007nd}, the relation between the thermal expectation value of the stress-energy tensor $T_{\mu\nu}$ and the metric fluctuations due to the trailing string has been worked out in detail.  Here we present a brief summary.  Let us write the spacetime metric as $ds^2 = G_{\mu\nu} dx^\mu dx^\nu$, with $G_{\mu\nu}=G^{(0)}_{\mu\nu}+ h_{\mu\nu}$, where $G^{(0)}_{\mu\nu}$ is the AdS${}_5$-Schwarzschild metric
\[
	ds_{(0)}^2 = G^{(0)}_{\mu\nu} dx^\mu dx^{\nu} = \frac{L^2}{z^2}\left[-g(z)dt^2+\sum_i dx_i^2 + g(z)^{-1} dz^2\right]
	\qquad g(z)=1-\frac{z^4}{z_0^4}\,,
\]
and $h_{\mu\nu}$ is the first order correction to it in response to the trailing string.  The temperature $T$ of the field theory is related to the location $z_0$ of the black hole horizon through $z_0 = 1/\pi T$.  The stress-energy tensor of the boundary theory is proportional to the fourth order coefficient in the expansion of the metric fluctuations $h_{\mu\nu}(z)$ at small $z$.  The stress-energy tensor is traceless, as expected for a conformal theory, but it is not conserved.  The non-conservation simply means that energy, as well as momentum in the direction of the quark's motion, are deposited into the thermal plasma at a constant rate \cite{Friess:2006fk}.  This rate reproduces the drag force as computed in \cite{Herzog:2006gh,Gubser:2006bz}.

In order to obtain the stress-energy tensor explicitly, one has to solve the linearized Einstein equations for the metric fluctuations $h_{\mu\nu}$ sourced by the trailing string.  These equations take the form
\begin{equation}
\label{E:hEOM}
	\mathcal{D}^{\mu\nu\rho\sigma}h_{\rho\sigma} = J^{\mu\nu}\,,
\end{equation}
where $\mathcal{D}^{\mu\nu\rho\sigma}$ is a second order differential operator (a Lichnerowicz operator) and $J^{\mu\nu}$ is the source term generated by the string.  These equations can be reduced to five decoupled second order differential equations written in terms of gauge-invariant quantities called ``master fields'' \cite{Gubser:2007nd}.  Because of parity considerations, two of these equations have no source terms.  The two corresponding master fields can be set to zero, and so they don't contribute to $h_{\mu\nu}$.  The remaining three master equations can be written in the form
\begin{equation}
\label{E:maindiffeq}
	\left[z^3 \partial_z z^{-3} g \partial_z -1+v^2 k_1^2\left(g^{-1}-1\right)+\tilde{k}^2 V_X(z\tilde{k}) \right]\tilde{k}^{-2}\Phi_X(z\tilde{k}) = J_X(z\tilde{k})\,,
\end{equation}
with $X=S$, $V$, and $T$ corresponding to ``scalar,'' ``vector,'' and ``tensor'' perturbations, respectively (all with even parity).  Also, $(k_1,k_2,k_3)$ are the variables conjugate to $(x_1,x_2,x_3)$ (recall that our metric signature is mostly plus), and $\tilde{k}^2 = k^2-v^2 k_1^2$.  In what follows we shall find it convenient to switch to the dimensionless variables $\vec{K} = z_0 \vec{k}$ and $\vec{X} = \vec{x}/z_0$, as defined in \eqref{KXDef}.  We will also use a convenient rescaling of the ``depth'' coordinate in AdS${}_5$, namely $Z = z \tilde{k}$.

The quantity ${\cal E}(\vec{K})$ appearing in \eqref{FTDef} can be obtained from a small $Z$ expansion of the scalar master field $\Phi_S$:
\begin{equation}
\label{E:PhiSQE}
	\Phi_S = \ldots - \tilde{K}^{-2}\left( - 12 \pi \mathcal{E}(\vec{K}) v K_1  + i 6 v^2 \right) Z^2 + \ldots\,,
\end{equation}
where $\tilde{K} = \sqrt{K_1^2(1-v^2) + K_\perp^2}$ and $K_\perp = \sqrt{K_2^2 + K_3^2}$ (see \cite{Gubser:2007nd} for details.)

The small $K$ asymptotics of $\mathcal{E}(\vec{K})$ may be obtained by expanding $\Phi_S$ in power series in $K$, $\Phi_S = \sum K^n \psi_n$, and solving the corresponding equation in \eqref{E:maindiffeq} perturbatively in small $K$ \cite{Friess:2006fk,Gubser:2007nd} (see equation \eqref{E:QElowk} below.)  The large $K$ asymptotics can be obtained in a similar manner as a power series in $1/K$, which may also be regarded as a series in positive powers of $T$.  This series was calculated up to order ${\cal O}(K^{-1})$ in \cite{Yarom:2007ni,Gubser:2007nd}.  In appendix~\ref{LARGEK} we rederive these results and extend them to order $\mathcal{O}(K^{-3})$.  The values of $\mathcal{E}(\vec{K})$ for intermediate values of $K$ have to be evaluated numerically.  We do this using the method developed in \cite{Friess:2006fk}.

With small $K$ and large $K$ asymptotics in hand, we consider a decomposition
 \eqn{Edecompose}{
  {\cal E}(\vec{K}) = {\cal E}_{\rm UV}(\vec{K},\mu_{\rm UV}) +
    {\cal E}_{\rm IR}(\vec{K},\mu_{\rm IR}) +
    {\cal E}_{\rm res}(\vec{K},\mu_{\rm UV},\mu_{\rm IR}) \,.
 }
Here ${\cal E}_{\rm UV}(\vec{K},\mu_{\rm UV})$ agrees with the large $K$ asymptotics up to and including terms of order $K^{-3}$; ${\cal E}_{\rm IR}(\vec{K},\mu_{\rm IR})$ almost agrees with the small $K$ asymptotics up to and including terms of order $K^0$ (see equation \eqref{ImperfectMatch} and the discussion following it for more precise details); and ${\cal E}_{\rm res}(\vec{K},\mu_{\rm UV},\mu_{\rm IR})$ is uniformly bounded and integrable.  The parameters $\mu_{\rm UV}$ and $\mu_{\rm IR}$ are pure numbers which can be adjusted to make the residual part ${\cal E}_{\rm res}(\vec{K})$ as small as possible.  The precise analytic forms of ${\cal E}_{\rm UV}(\vec{K},\mu_{\rm UV})$ and ${\cal E}_{\rm IR}(\vec{K},\mu_{\rm IR})$ will be explained in sections~\ref{NEAR} and~\ref{FAR}, respectively.

Ideally, ${\cal E}_{\rm UV}(\vec{K},\mu_{\rm UV})$ and ${\cal E}_{\rm IR}(\vec{K},\mu_{\rm IR})$ should admit analytic Fourier transforms.  We didn't quite realize this goal: the Fourier transform of ${\cal E}_{\rm IR}(\vec{K},\mu_{\rm IR})$ is left in the form of a one-dimensional Fourier integral which must be performed numerically.  The Fourier transform of ${\cal E}_{\rm res}(\vec{K},\mu_{\rm UV},\mu_{\rm IR})$ must also be performed numerically, via a three-dimensional fast Fourier transform.  In section~\ref{NUMERICAL} we report the results of numerics for three values of velocity: $v=0.25$, $v=0.58$, and $v=0.75$.  The second of these is only slightly larger than the speed of sound in the thermal plasma, $c_s = 1/\sqrt{3} \approx 0.577$.

\subsection{Near-field asymptotics}
\label{NEAR}

Using methods explained in section~\ref{REVIEW} and appendix~\ref{LARGEK}, one finds
 \eqn{HighKAsymptForm}{
  {\cal E}(\vec{K}) = {\cal E}^{(0)}_{{\rm UV}}(\vec{K}) +
    {\cal O}(K^{-5})\,,
 }
where
 \eqn{HighKAsympt}{
  {\cal E}^{(0)}_{\rm UV}(\vec{K}) &=
   -{K_1^2 v^2 (-1+v^2) + \tilde{K}^2 (2+v^2)
    \over 24 \tilde{K}} -
   {i K_1 v [2K_1^2 v^2 (-1+v^2) + \tilde{K}^2 (-5+11v^2)] \over
     18\pi \tilde{K}^4}  \cr
   &\qquad\qquad{} + {3K_1^4 v^4 (-1+v^2) + 7 \tilde{K}^4 (2+v^2) +
    K_1^2 \tilde{K}^2 v^2 (-1+10v^2) \over 24 \tilde{K}^7}\,,
 }
where, as before, $\tilde{K} = \sqrt{K_1^2 (1-v^2) + K_\perp^2}$ and $K_\perp = \sqrt{K_2^2+K_3^2}$.  The difference ${\cal E}(\vec{K}) - {\cal E}^{(0)}_{{\rm UV}}(\vec{K})$ is small at large $K$ but large at small $K$ because of the inverse powers of $\tilde{K}$ that appear in \eqref{HighKAsympt}.  This is bad because our eventual aim is to find an analytic approximation to ${\cal E}(\vec{K})$ that is good both in the UV and the IR.  The bad IR behavior of \eno{HighKAsympt} can be cured by using the identity
 \eqn{KtildeReplace}{
  {1 \over \tilde{K}^n} = {1 \over
   (\tilde{K}^2 + \mu_{\rm UV}^2)^{n/2}}
    \left[ 1 - {\mu_{\rm UV}^2 \over \tilde{K}^2 + \mu_{\rm UV}^2}
      \right]^{-n/2}
 }
and expanding the quantity in square brackets to just enough terms to keep the ${\cal O}(K^{-5})$ accuracy that ${\cal E}^{(0)}_{{\rm UV}}(\vec{K})$ possesses in the first place.  Through this procedure one obtains
\begin{equation}
\label{EUVreg}
    \mathcal{E}_{\rm UV}(\vec{K},\mu_{\rm UV})=
    -\frac{\left( 2 + v^2 \right) \,{\sqrt{\tilde{K}^2 + \mu_{\rm UV}^2}}}{24}
    +
    \frac{-2\,K_1^2\,v^2\,\left( -1 + v^2 \right)  + \left( 2 + v^2 \right) \,\mu_{\rm UV}^2}{48\,{\sqrt{\tilde{K}^2 + \mu_{\rm UV}^2}}}
    +
    \ldots \,,
\end{equation}
where the omitted terms are polynomials in $K_1$, $\mu_{\rm UV}$, and $v$ times negative powers of $\sqrt{\tilde{K}^2+\mu_{\rm UV}^2}$.  These terms are straightforward to work out, but their precise form is long and not very enlightening.

The Fourier transform of ${\cal E}_{\rm UV}(\vec{K},\mu_{\rm UV})$, as well as several other Fourier transforms required in later sections, can be worked out starting from
 \eqn{FTmain}{
  \int {d^d K \over (2\pi)^d} {e^{i \vec{K} \cdot \vec{X}} \over
        (K^2 + \mu^2)^n}
    = {2 \over (4\pi)^{d/2} \Gamma(n)} \left( {X \over 2\mu}
       \right)^{n-d/2} {\bf K}_{n-d/2}(\mu X)
 }
and taking appropriate derivatives of it.  Here $\BesselK{\nu}(z)$ is a modified Bessel function of the second kind. For the case at hand we use \eno{FTmain} with $d=3$ and $K \to \tilde{K}$.

The Fourier transform of ${\cal E}_{\rm UV}(\vec{K},\mu_{\rm UV})$ is
\begin{multline}
\label{E:EUVregreal}
	\mathcal{E}_{\rm UV}(\vec{X},\mu_{\rm UV})=
	  -\frac{\mu_{\rm UV}^2}{96 {\pi }^2
	   {\left( 1 - v^2 \right) }^{\frac{3}{2}} {\tilde{X} }^4}
        \left( \mu_{\rm UV}  \tilde{X}  \Big( 2 v^2 X_1^2 + \left( -2 + v^2 + v^4 \right)  {\tilde{X} }^2 \right)
        \BesselK{1}(\mu_{\rm UV} \tilde{X})
    \\+ 4 \left( 2 v^2 X_1^2 + \left( -1 + v^4 \right)  {\tilde{X} }^2 \right)
         \BesselK{2}(\mu_{\rm UV}  \tilde{X} ) \Big) + \ldots \,,
\end{multline}
where we have defined $\tilde{X}=\sqrt{X_1^2 / (1-v^2)+X_{\bot}^2}$.  The terms written explicitly in (\ref{E:EUVregreal}) correspond to those appearing explicitly in \eqref{EUVreg}.  Additional terms have similar forms, and it would not be very illuminating to write them out explicitly here.

\subsection{Far-field asymptotics}
\label{FAR}

As described at the beginning of this section, the IR asymptotics of the energy density may be obtained by solving equation \eqref{E:maindiffeq} perturbatively in small $K$.  One obtains
\begin{equation}
\label{E:QElowk}
	\mathcal{E}(\vec{K}) =
	-\frac{3 i K_1 v
   	\left(v^2+1\right)}{2 \pi(K_1^2(1-3v^2)+K_{\bot}^2) }
	+
	\frac{3 K_1^2 v^2 \left(K_{\bot}^2(2+v^2)+2 K_1^2(1+v^2)\right)}{2 \pi {(K_1^2(1-3v^2)+K_{\bot}^2)}^2 }
	+
	\mathcal{O}(K)\,.
\end{equation}
A problem with \eqref{E:QElowk} is that the Fourier transform of negative powers of $K_1^2(1-3v^2)+K_{\bot}^2$ is not well defined
when $v^2>1/3$ due to the poles which appear on the real axis.  As suggested in \cite{Friess:2006fk}, the quantity
\eqn{EKResummedAgain}{
  {\cal E}^{\rm (resummed)}_{\rm IR}(\vec{K}) = {{\cal A}(K_1)\over K_\perp^2 + {m(K_1)}^2}}
with
 \eqn{AmDef}{
  {\cal A}(K_1) &= - {3 i v K_1 (1+v^2)^2\over 2 \pi \left(1 + v^2 - i K_1 v (2+v^2) \right)}  \cr
  {m(K_1)}^2 &= -{(3v^2 - 1) (1+v^2) K_1^2 + 2 i K_1^3 v (1-v^2)
    \over 1 + v^2 - i K_1 v (2+v^2)}\,
 }
is a more uniform approximation to ${\cal E}(\vec{K})$ for small $\vec{K}$ than the terms shown explicitly in \eqref{E:QElowk}.  The expressions (\ref{E:QElowk}) and~\eno{EKResummedAgain} agree up to ${\cal O}(K)$ corrections, but the denominator of \eno{EKResummedAgain} shifts the poles away from the real axis.  We describe this modification of the denominator as a ``resummation'' of the series \eqref{E:QElowk} because a single rational expression, \eno{EKResummedAgain}, includes both terms in \eqref{E:QElowk}.\footnote{As will be discussed in section~\ref{CONCLUSIONS}, this resummation can be motivated physically in terms of passing from inviscid to viscous hydrodynamics.}

The problem now is that an analytic Fourier transform of $\mathcal{E}^{(\rm resummed)}_{\rm IR}(\vec{K})$ to real space is unavailable (as far as we know) because of the cubic terms in the denominator.  However, the integration over $\vec{K}_\perp = (K_2,K_3)$ may be done analytically using \eno{FTmain} with $d=2$ and $n=1$.  The result is
 \eqn{EXInt}{
  {\cal E}^{\rm (resummed)}_{\rm IR}(\vec{X}) = \int {d^3K\over (2 \pi)^3}
     e^{i \vec{K}\cdot \vec{X}} {{\cal A}(K_1) \over K_\perp^2 + {m(K_1)}^2}
   = \int_{-\infty}^\infty {dK_1\over 2 \pi}
     e^{i K_1 X_1} {{\cal A}(K_1)\over 2 \pi} {\bf K}_0(m(K_1) X_\perp)
 }
where $m(K_1) = \sqrt{{m(K_1)}^2}$ is chosen so that $\Re\{m(K_1)\}>0$.  The last integral in \eno{EXInt} does not seem to be expressible in closed form.  However, for $X_\perp > 0$, the integrand has exponential decay for large $\abs{K_1}$, and it is continuous.  So it can be handled reliably with numerical methods, i.e.~a one-dimensional FFT\@.

The large $K$ behavior of ${\cal E}^{\rm (resummed)}_{\rm IR}(\vec{K})$ is ${\cal O}(1/K^2)$, so it brings back problems with the UV region that ${\cal E}_{\rm UV}(\vec{K},\mu_{\rm UV})$ was supposed to have fixed.  Let us therefore consider the following corrected form:
 \eqn{EIR}{
  \mathcal{E}_{\rm IR}(\vec{K},\mu_{\rm IR}) =
  	\frac{\mathcal{A}(K_1)}{K_{\bot}^2+m(K_1)^2} - \frac{\mathcal{A}(K_1)}{K_{\bot}^2+m(K_1)^2-\frac{i K_1 v \mu_{\rm IR}}{(1+v^2)-i K_1 v (2+v^2)}}\,.
 }
The Fourier transform of the second term in \eno{EIR} is similar to \eno{EXInt}.  At small $K$ one finds
 \eqn{ImperfectMatch}{
  \mathcal{E}(\vec{K})-\mathcal{E}_{\rm IR}(\vec{K},\mu_{\rm IR}) = -
     \frac{3(1+v^2)^2}{2\pi \mu_{\rm IR}^2}+\mathcal{O}(K) \,.
 }
Better would have been a form for ${\cal E}_{\rm IR}(\vec{K},\mu_{\rm IR})$ which matched ${\cal E}(\vec{K})$ up to ${\cal O}(K)$ corrections, with no ${\cal O}(K^0)$ term as in \eno{ImperfectMatch}: for instance, some exponential suppression of this term could have been arranged.  But, because the deviation of the subtracted quantity
$\mathcal{E}_{\rm IR}(\vec{K},\mu_{\rm IR})$ from
$\mathcal{E}(\vec{K})$ is finite at small $K$, it does not introduce divergences at small values of
the momenta and its Fourier transform can be easily dealt
with numerically.

\subsection{Numerical results}
\label{NUMERICAL}

We obtained ${\cal E}(\vec{K})$ numerically using minor improvements over the code used for \cite{Friess:2006fk}.  We chose $\mu_{\rm UV}=\mu_{\rm IR}=1$.  A three-dimensional FFT of $\mathcal{E}_{\rm res}=\mathcal{E}-\mathcal{E}_{\rm UV}-\mathcal{E}_{\rm IR}$ was performed on a grid with $128$ points on each side and with each component $K_i$ running approximately from $-10$ to $10$.  One-dimensional FFT's of $\mathcal{E}_{\rm IR}$ were carried out on grids with approximately $3000$ points, with $K_1$ again running approximately from $-10$ to $10$.  Edge effects can be expected in position space near the boundary of an FFT\@.  To smooth them out we multiplied the position space results of FFT's by a $C^1$ approximation of the top-hat function, i.e.~a function which is $0$ outside a large cube, $1$ inside a smaller cube, and has one continuous derivative everywhere.

\begin{figure}
  \centerline{\includegraphics[width=7in]{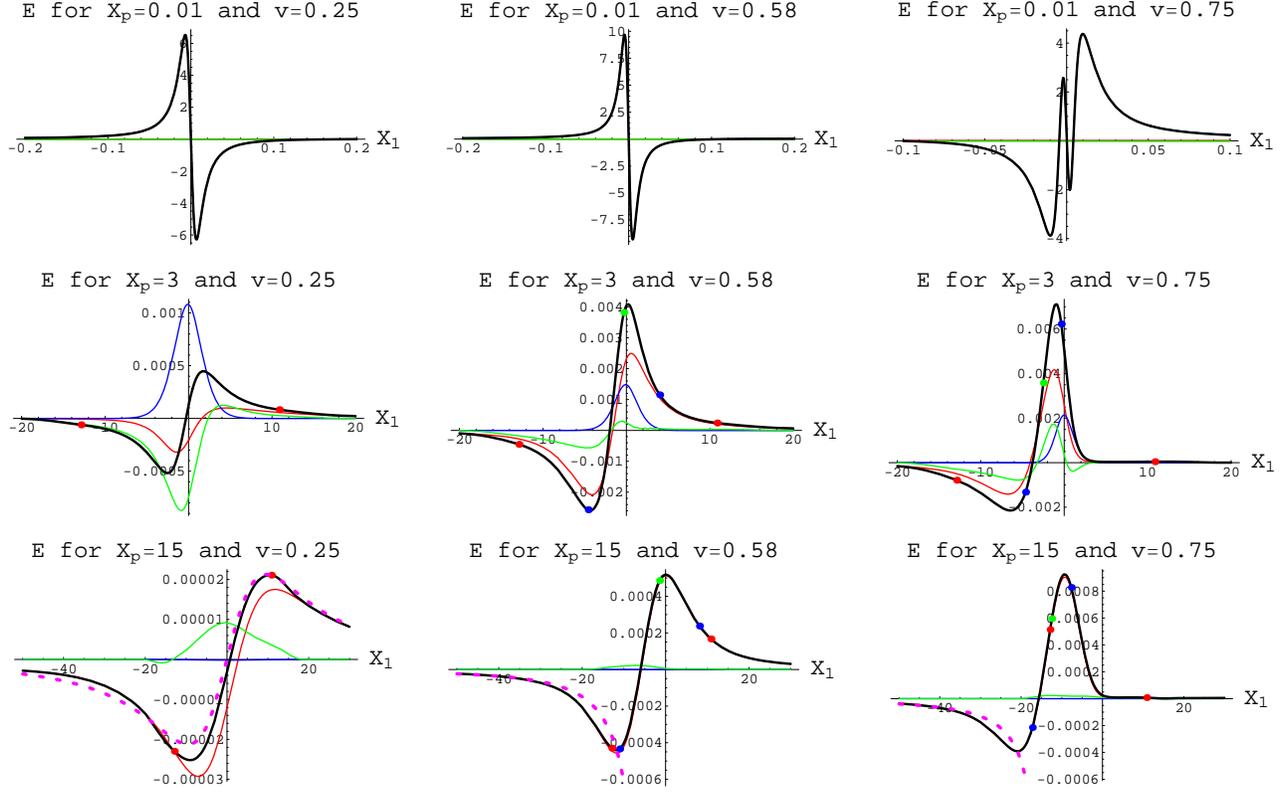}}
  \caption{(Color online.)  $E \equiv {\cal E}-{\cal E}_{\rm Coulomb}$ as a function of dimensionless position $X_1$ for various values of offset $X_p = X_\perp$ from the quark and velocity $v$ of the quark.  In each plot, the black curve is $E$.  The red curve is ${\cal E}_{\rm IR}$.  The green curve is ${\cal E}_{\rm res}$ with $C^1$ smoothing (see the main text).  The blue curve is ${\cal E}_{\rm UV}-{\cal E}_{\rm Coulomb}$.  The dotted purple curve is from inviscid linearized hydrodynamics (see the discussion in section~\ref{CONCLUSIONS}).  The green dot corresponds to the Mach angle.  The blue dots show a width typical of hydrodynamical broadening (see the main text).  The red dots show the points where the smoothing of ${\cal E}_{\rm res}$ starts.  The quark is at $X_1 = X_\perp = 0$.}\label{ComparisonColumns}
\end{figure}

In figure~\ref{ComparisonColumns} we plot the net energy density $E \equiv {\cal E} - {\cal E}_{\rm Coulomb}$ as a function of $X_1$ at various velocities and various values of $X_\perp$.
To get a feel for the scales involved, note that if $T = {1 \over \pi}\,{\rm GeV} \approx 318\,{\rm MeV}$, then $X=1$ corresponds to a distance $1\,{\rm GeV}^{-1} \approx 0.2\,{\rm fm}$ from the quark.

The lower left plot in figure~\ref{ComparisonColumns} shows that our infrared subtractions are not as accurate multiplicatively for $v < 1/\sqrt{3}$ as they are for $v > 1/\sqrt{3}$.  Probably a better subtraction scheme could be contrived for $v < 1/\sqrt{3}$ starting from the quantity ${\cal E}_{\rm IR}^{\rm (inviscid)}$ discussed in section~\ref{CONCLUSIONS}.  Note however that the magnitude of $E$ is very small in the region in question.  The middle left plot in figure~\ref{ComparisonColumns} shows that the asymptotic form ${\cal E}_{\rm UV}(\vec{X},\mu_{\rm UV})$ deviates strongly from the full expression for $E$ on intermediate length scales when $v=0.25$.  This is not particularly alarming: the scales in question are far from the ultraviolet regime.  However, one might hope by cleverer choice of asymptotic forms (perhaps a more felicitous choice of $\mu_{\rm UV}$ and $\mu_{\rm IR}$) to obtain a more uniformly accurate approximation to $E(\vec{X})$.

\begin{figure}
  \centerline{\includegraphics[width=4in]{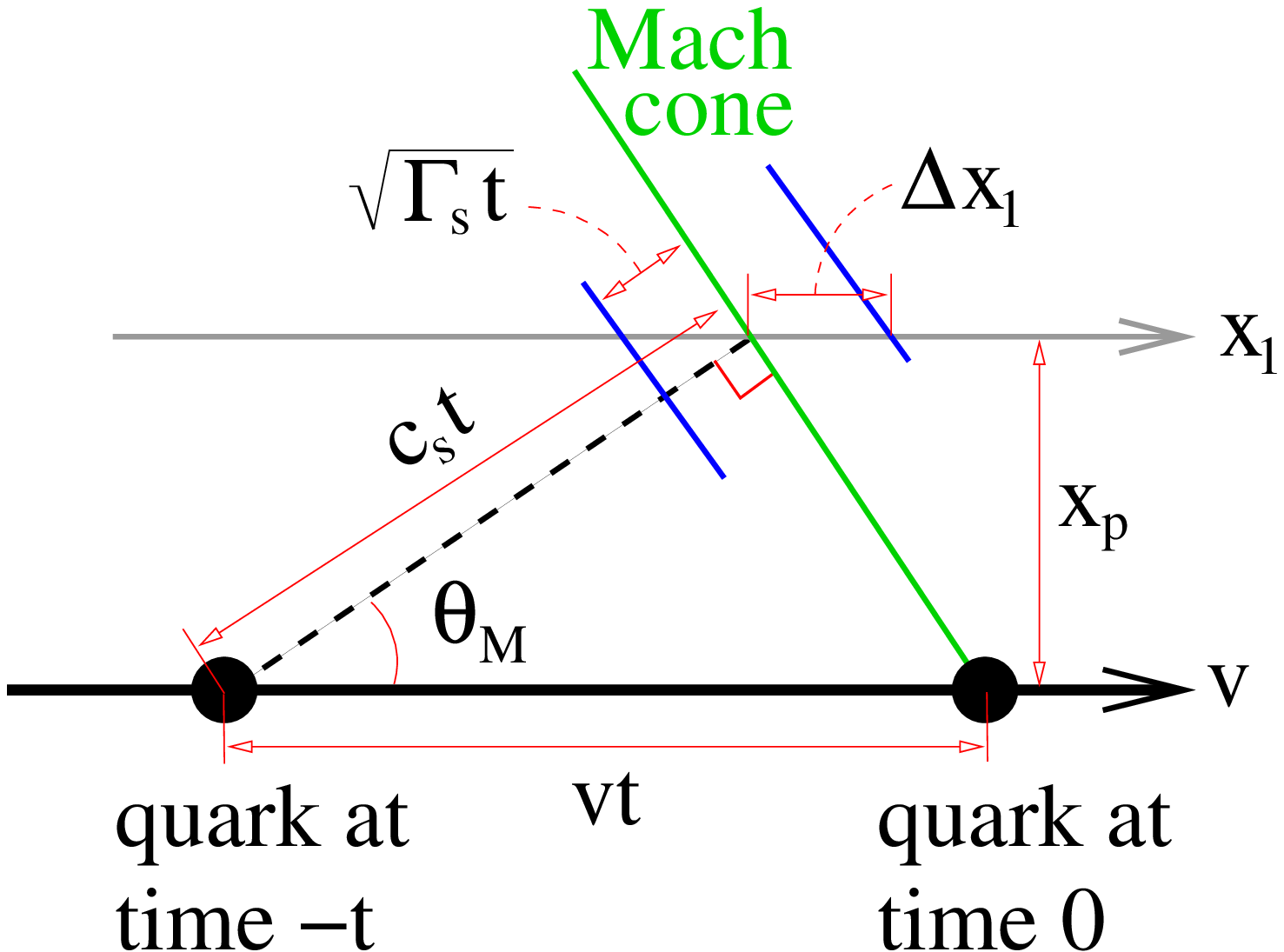}}
  \caption{(Color online.)  An estimate of hydrodynamical broadening of the Mach cone.  The gray line, offset by a distance $x_p = x_\perp$ from the quark's trajectory, is the axis along which we plot $E \equiv {\cal E}-{\cal E}_{\rm Coulomb}$ in figure~\ref{ComparisonColumns}.  The green dot in that figure is at the intersection of the gray and green lines in this one; likewise the blue dots are at intersections of the gray line with the blue lines.}\label{BroadenedMach}
 \end{figure}

For supersonic velocities, one can compare the structure associated with a Mach cone to the scale of hydrodynamical broadening, $\Delta x = \sqrt{\Gamma_s t}$.  Here $\Gamma_s$ is the diffusion length for sound waves.  We plug in $t={x_\perp \over c_s} \csc\theta_M$, where $\theta_M = \arccos {c_s \over v}$ is the Mach angle, because this is the time it takes for sound to travel from the quark to a point on the Mach cone a transverse distance $x_\perp$ from the axis of the quark's motion: see figure~\ref{BroadenedMach}.
The sound waves propagate at the Mach angle, whereas our plots in figure~\ref{ComparisonColumns} are along the $X_1$ direction.  Thus the expected broadening in $x_1$ is
 \eqn{xOneBroadening}{
  \Delta x_1 = \Delta x \, \sec\theta_M =
    \sqrt{\Gamma_s {X_\perp \over c_s} \csc\theta_M} \,
     \sec\theta_M \,.
 }
Using $c_s = 1/\sqrt{3}$, $\Gamma_s = 1/3\pi T$, and $X_1 = \pi T x_1$, one finds
 \eqn{xOneAgain}{
  \Delta X_1 = \pi T \Delta x_1 = {\sqrt{3v^3 X_\perp} \over
   \sqrt[4]{3v^2-1}} \,.
 }

 \begin{figure}
  \centerline{\includegraphics[width=4in]{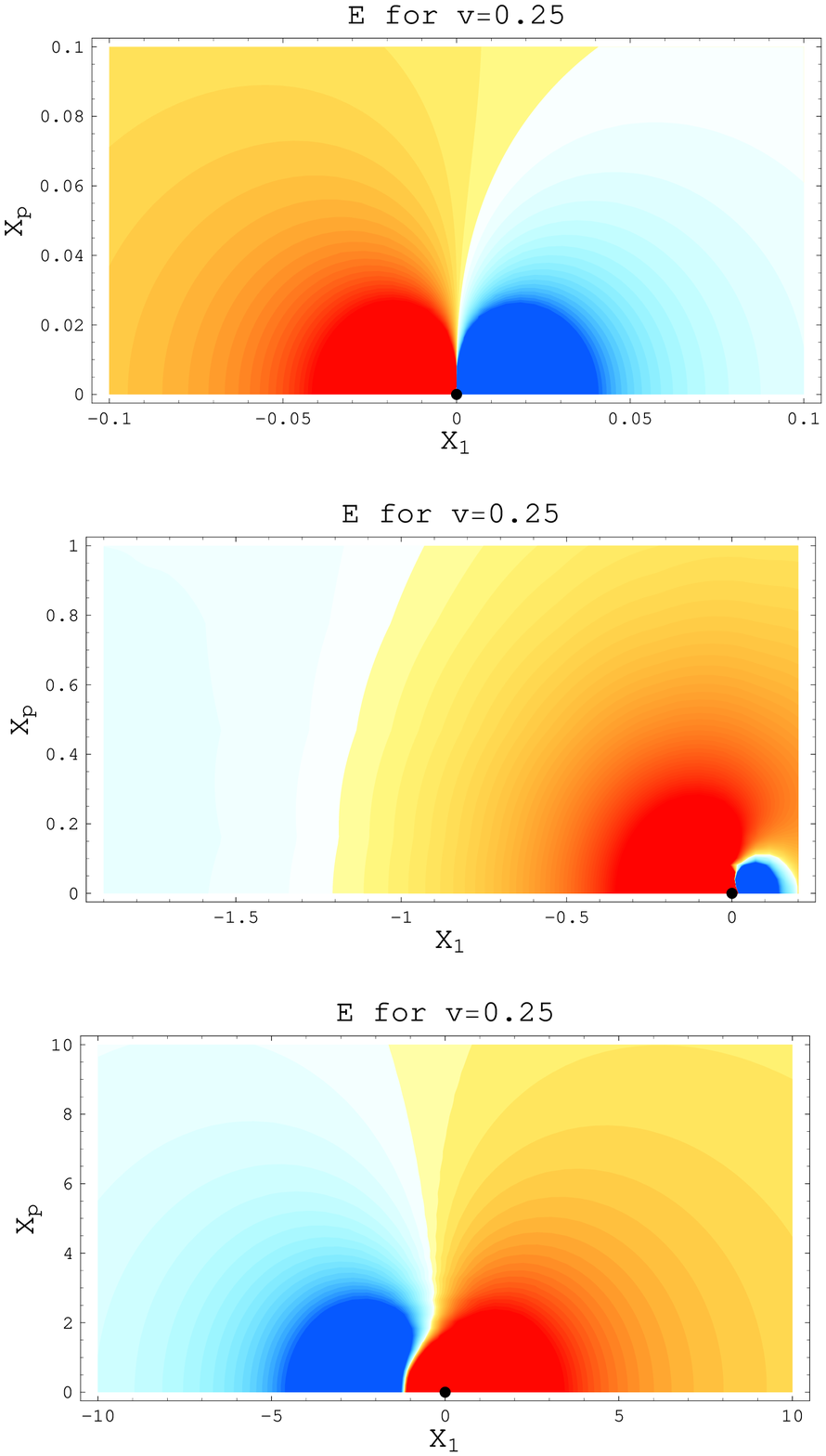}}
  \caption{(Color online.)  Contour plots of $E \equiv {\cal E}-{\cal E}_{\rm Coulomb}$ as a function of dimensionless position coordinates $X_1$ and $X_p=X_\perp$, for $v=0.25$.  Note that the top plot shows the smallest structures while the bottom plot shows the largest.  Orange and red regions correspond to $E>0$; white regions correspond to $E\approx 0$; and blue regions correspond to $E < 0$.  The energy density of the thermal bath is not included in $E$.  The three-dimensional energy density profile is axially symmetric around the $X_1$ axis.  The black dot is the position of the quark: $X_1=X_\perp = 0$.}\label{ContourColumn25}
 \end{figure}
 \begin{figure}
  \centerline{\includegraphics[width=4in]{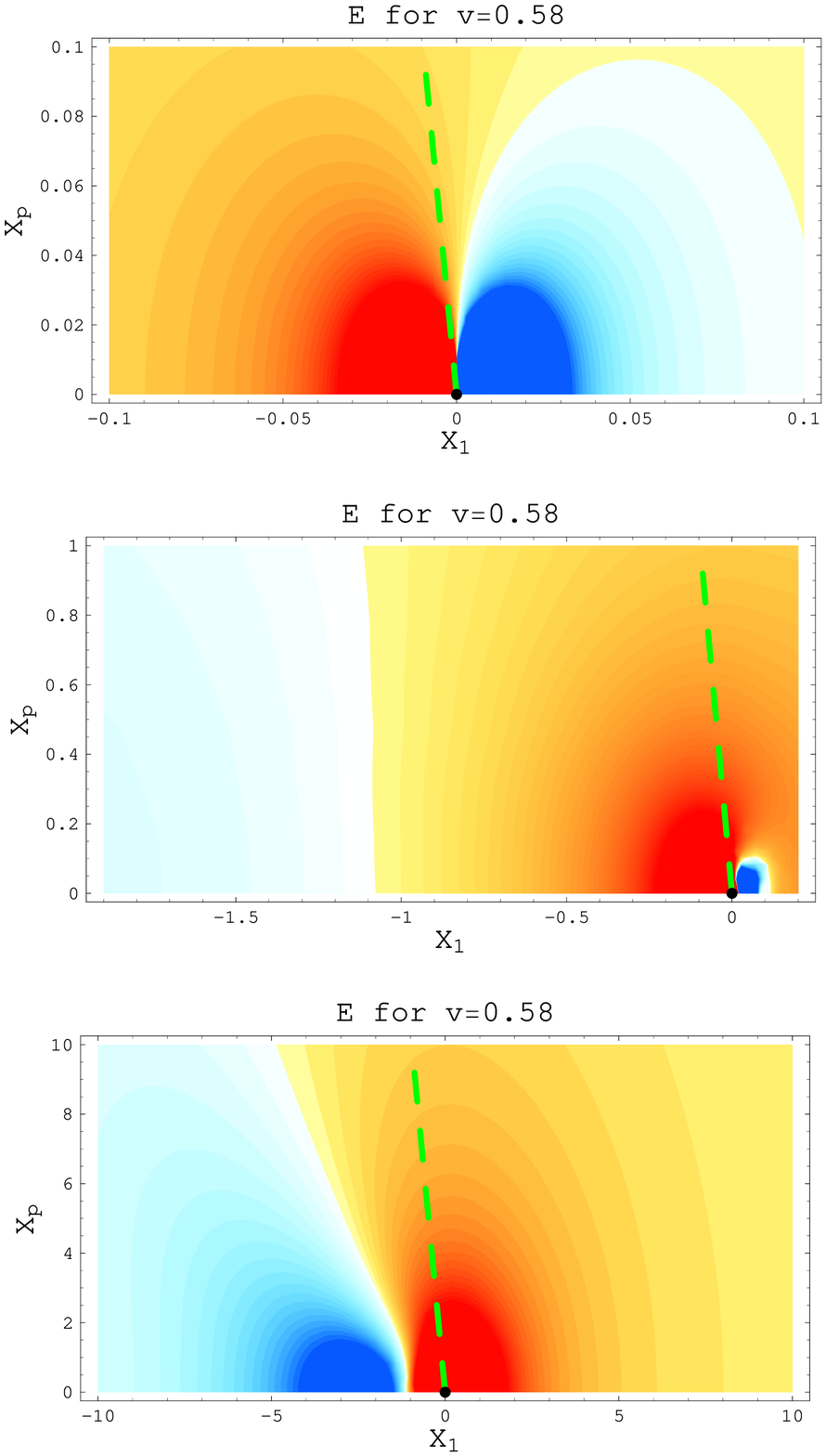}}
  \caption{(Color online.)  Contour plots of $E \equiv {\cal E}-{\cal E}_{\rm Coulomb}$ as a function of dimensionless position coordinates $X_1$ and $X_p=X_\perp$, for $v=0.58$.  Note that the top plot shows the smallest structures while the bottom plot shows the largest.  Orange and red regions correspond to $E>0$; white regions correspond to $E\approx 0$; and blue regions correspond to $E < 0$.  The energy density of the thermal bath is not included in $E$.  The three-dimensional energy density profile is axially symmetric around the $X_1$ axis.  The black dot is the position of the quark: $X_1=X_\perp = 0$.  The dashed green line shows the Mach cone.}\label{ContourColumn58}
 \end{figure}
 \begin{figure}
  \centerline{\includegraphics[width=4in]{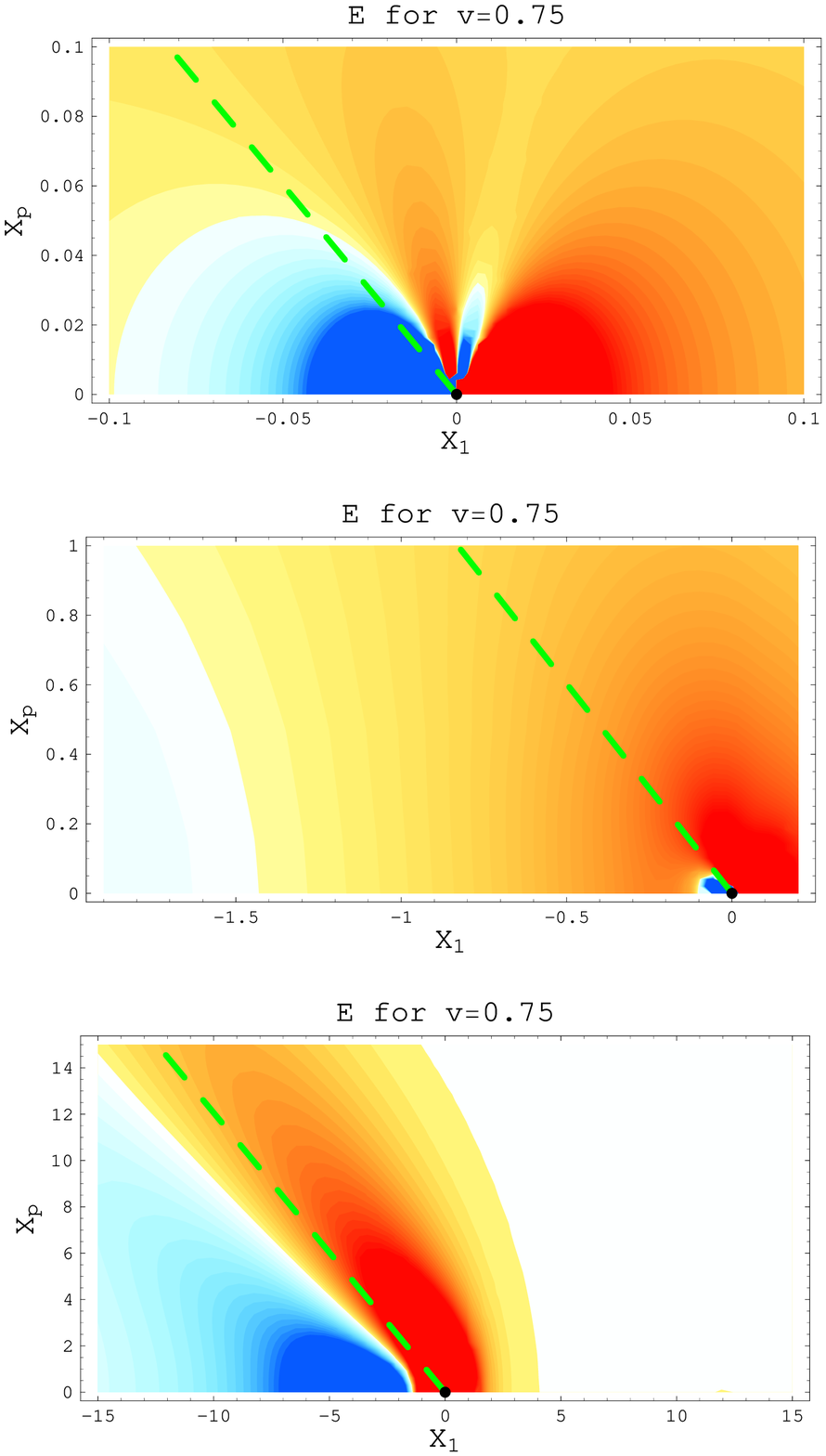}}
  \caption{(Color online.)  Contour plots of $E \equiv {\cal E}-{\cal E}_{\rm Coulomb}$ as a function of dimensionless position coordinates $X_1$ and $X_p=X_\perp$, for $v=0.75$.  Note that the top plot shows the smallest structures while the bottom plot shows the largest.  Orange and red regions correspond to $E>0$; white regions correspond to $E\approx 0$; and blue regions correspond to $E < 0$.  The energy density of the thermal bath is not included in $E$.  The three-dimensional energy density profile is axially symmetric around the $X_1$ axis.  The black dot is the position of the quark: $X_1=X_\perp = 0$.  The dashed green line shows the Mach cone.}\label{ContourColumn75}
 \end{figure}

Figures~\ref{ContourColumn25}, \ref{ContourColumn58}, and~\ref{ContourColumn75} show contour plots of the net energy density $E \equiv {\cal E} - {\cal E}_{\rm Coulomb}$ for various values of the quark velocity.  The structures exhibited in $E \equiv {\cal E}-{\cal E}_{\rm Coulomb}$ are notably scale-dependent.  Observe for example from figure~\ref{ContourColumn25} that at $v=0.25$, $E>0$ for small negative and large positive $X_1$ and $E<0$ for small positive and large negative $X_1$.  Recall that if $T = {1 \over \pi} \, {\rm GeV} \approx 318\,{\rm MeV}$, then $X=1$ corresponds to a distance $1\,{\rm GeV}^{-1} \approx 0.2\,{\rm fm}$ from the quark.

Note that the double-peaked feature in the upper right plot in figure~\ref{ComparisonColumns} corresponds to the structure near the origin of the top plot in figure~\ref{ContourColumn75}.  As observed in \cite{Gubser:2007nd,Yarom:2007ni}, in three dimensions this structure is a forward lobe and a backward-leaning cone of energy over-density (as usual, relative to the sum of the energy density of the moving quark and the thermal bath in the absence of interaction between the two), with regions of under-density in complementary regions.

In the large scale plots one sees that the net energy density falls quickly to zero for positive $X_1$ when $v > 1/\sqrt{3}$.  We suspect that ${\cal E}$ (not $E$) falls exponentially in this direction.  For negative $X_1$ and fixed $X_\perp$, $E$ decays as $1/X_1^2$ (see equation (\ref{HydroXfast})).

\section{Conclusions}
\label{CONCLUSIONS}

Through a combination of analytic and numerical methods, we have calculated the position space energy density of a quark moving through a thermal state of ${\cal N}=4$ plasma.  More precisely, we considered the net energy density, obtained by subtracting away the constant contribution from the thermal bath as well as the field the quark would have generated in the absence of the bath.  This net energy density provides some gauge-invariant information about energy loss.  It has interesting structure on multiple scales, from sizes $\sim 0.01\,{\rm fm}$ to $\sim 2\,{\rm fm}$ if we take $T = 318\,{\rm MeV}$.  By comparing asymptotic forms (analytic or semi-analytic) to numerical results we can be reasonably confident that we have accurately characterized these structures.  The power of AdS/CFT is that all length scales can be treated at once.

The main features of our results can be understood in terms of analytic approximations.  At large distances and supersonic velocities, the Mach cone can be described by the small momentum asymptotics of the energy density,
 \eqn{Eapprox}{
  {\cal E}(\vec{K}) \sim {1 \over K^2 - 3v^2 K_1^2 - i v K^2 K_1}
   \,,
 }
as found in \cite{Friess:2006fk}.  (We have excluded from the denominator in \eno{Eapprox} some cubic terms that vanish on the Mach cone.)  If the cubic term in \eno{Eapprox} were dropped, then for $v > 1/\sqrt{3}$ there would be poles on the $K_1$ axis corresponding to zeroes of $K^2 - 3v^2 K_1^2$.  An integration prescription is required to pass these poles, and the physical one is given by the $- i v K^2 K_1$ term.  Carrying out this contour integral one finds a Mach cone structure in real space, as could have been expected from linearized hydrodynamics.

The relationship of \eno{Eapprox} to linearized hydrodynamics is worth understanding a bit better.  According to linearized hydrodynamics, the energy density obeys an equation of the form
 \eqn{ComovingHydro}{
  \left[ {\partial^2 \over \partial t^2} - {\partial^2 \over
    \partial \vec{x}^2} (c_s^2 + \Gamma_s \partial_t) \right]
     \epsilon = \hbox{sources} \,.
 }
(See for example \cite{Casalderrey-Solana:2006sq}.)  Comparing the Fourier transform of \eno{ComovingHydro} to the denominator of \eno{Eapprox}, and recalling that all dependence of $\epsilon$ on time and $x_1$ is in terms of $x_1-vt$, one finds perfect agreement if $c_s^2 = 1/3$ and $\Gamma_s = 1/3\pi T$.  In general, $\Gamma_s = 4 \eta/ 3 s T$, so this provides an independent check of the result $\eta/s = 1/4\pi$ \cite{Policastro:2001yc}.  Evidently, the cubic term in the denominator of \eno{Eapprox} arises from viscosity.  If it is removed, corresponding to sending $\eta \to 0$, then the energy density becomes singular on the Mach cone.  More explicitly: the inviscid linearized hydrodynamics approximation to \eno{EKResummedAgain} is
 \eqn{HydroK}{
  {\cal E}_{\rm IR}^{\rm (inviscid)}(\vec{K}) =
    -{3iv (1+v^2) \over 2\pi} {K_1 \over K_1^2 (1-3v^2) + K_\perp^2
      - i\varepsilon K_1} \,,
 }
where the infinitesimal positive quantity $\varepsilon$ provides the correct pole passing prescription. One easily finds
 \eqn{HydroX}{
  {\cal E}_{\rm IR}^{\rm (inviscid)}(\vec{X}) =
    {3v(1+v^2) \over 8\pi^2} {X_1 \over
     (X_1^2 + (1-3v^2) X_\perp^2)^{3/2}}
 }
for $v < 1/\sqrt{3}$, and
 \eqn{HydroXfast}{
  {\cal E}_{\rm IR}^{\rm (inviscid)}(\vec{X}) =
    {3v(1+v^2) \over 4\pi^2} {X_1 \over
     (X_1^2 + (1-3v^2) X_\perp^2)^{3/2}} \theta(-X_1-
       X_\perp \sqrt{3v^2-1})
 }
for $v > 1/\sqrt{3}$.  The functions \eno{HydroX} and \eno{HydroXfast} are shown in the large scale plots in figure~\ref{ComparisonColumns}.  The result \eno{HydroXfast} can be supplemented by a singular distribution of positive energy right on the Mach cone, but it is best regarded as an asymptotic form for large negative $X_1$.

The charm of the AdS/CFT calculation is that it interpolates smoothly all the way from \eno{Eapprox} down to the Coulombic near-field of the quark, for which $\epsilon \sim 1/x^4$ in the quark's rest frame.  A good understanding of the structure observed at scales $x \ll 1/T$ can already be obtained from considering just the analytic $\mathcal{O}(T^2)$ correction to the Coulombic field \cite{Gubser:2007nd,Yarom:2007ni}.  In particular, the over-density of energy ahead of the quark for $v = 0.25$ and $v = 0.58$ and the more complicated structure near the quark for $v=0.75$ were predicted by that work.  These non-trivial structures illustrate the need (already well recognized) to supplement hydrodynamics with a physical prescription of how energy is deposited near the quark. The next order $\mathcal{O}(T^4)$ corrections have been calculated in appendix~\ref{LARGEK}.

Extensions of the analytical computations in appendix~\ref{LARGEK}, as well as numerical treatments of other components of the stress tensor, are underway.  When complete, perhaps they will provide an interesting alternative perspective to perturbative intuitions about energy loss mechanisms in QCD\@.  Caution is appropriate when comparing AdS/CFT calculations to QCD in this context because ${\cal N}=4$ SYM is conformal, and therefore equally strongly coupled at the smallest and largest of scales.

\section*{Acknowledgments}

The work of S.~Gubser was supported in part by the Department of Energy under Grant No.\ DE-FG02-91ER40671, and by the Sloan Foundation. A.~Yarom is supported in part by the German Science Foundation and by the Minerva foundation. S.~Gubser and S.~Pufu thank J.~Friess and F.~Rocha for collaboration on earlier efforts to obtain the position space form of the energy density.  We thank K.~Rajagopal, D.~Teaney, and L.~Yaffe for useful discussions, P.~Chesler for correspondence and M.~Haack for comments on the manuscript.

\clearpage
\appendix

\section{Large momentum asymptotics}
\label{LARGEK}
The starting point for computing the large momentum asymptotics of the energy density ${\cal E}(\vec{K})$ is the scalar master equation in \eqref{E:maindiffeq}, with the potential
\begin{equation}
\label{E:VS}
	V_S =
		\frac{12\,\left( 2\,Z^4\,\left( 1 + {\alpha }^2 \right)  +
		2\,{K}^4\,\left( 1 + {\alpha }^2 \right)  +
		Z^2\,\left( -12 + {K}^4\,{\left( 1 + {\alpha }^2 \right) }^2 \right)  \right) }
		{{\left( 6\,Z^2 + {K}^4\,\left( 1 + {\alpha }^2 \right)  \right) }^2}
\end{equation}
and source term
 \eqn{JSDef}{
	J_S &=
	\frac{e^{-i K_1 \xi(Z)}}{Z\,
        {\left( 6 Z^2 + {K}^4 \left( 1 + {\alpha }^2 \right)  \right) }^2}
    \Bigg(
        v^2 \left( 1 + {\alpha }^2 \right)
        \Big( -54 Z^6 \alpha  + 3 {K}^8 \alpha \left( 1 + {\alpha }^2 \right)  \cr
     &\qquad\qquad{}-
        6i  Z^5 {K}^2 \left( 2 + 11 {\alpha }^2 \right)  +
        3 Z^4 {K}^4 \alpha  \left( 5 + 11 {\alpha }^2 \right)  +
        3i  Z^3 {K}^6
        \left( 2 + 5 {\alpha }^2 + 3 {\alpha }^4 \right)  \cr
     &\qquad\qquad\qquad\qquad\qquad\qquad{}+
       Z^2 {K}^4 \alpha  \left( 90 +
          {K}^4 \left( -2 + {\alpha }^2 \right)  \left( 1 + {\alpha }^2 \right)  \right)
           \Big)\cr\noalign{\vskip3\jot}
     &{}- \alpha  \Big(  -162 i  Z^5{K}^2
        {\alpha }^3 + \left( 9 i  \right)  Z^3 {K}^6 {\alpha }^3
        \left( 1 + {\alpha }^2 \right)  +
       3 {K}^8 \left( -2 + {\alpha }^2 \right)  \left( 1 + {\alpha }^2 \right)  \cr
     &\qquad\qquad{}-
       18 Z^6 \left( -2 + 7 {\alpha }^2 \right)  +
       3 Z^4 {K}^4 \left( 2 + 7 {\alpha }^2 + 23 {\alpha }^4 \right) \cr
     &\qquad\qquad\qquad\qquad\qquad\qquad{}+
       Z^2 {K}^4 \left( -2 + {\alpha }^2 \right)
        \left( 90 + {K}^4 \left( -2 + {\alpha }^2 \right)
           \left( 1 + {\alpha }^2 \right)  \right)  \Big)
    \Bigg)\,,
}
as constructed in \cite{Gubser:2007nd}. Here,
 \eqn{xiDef}{
  \xi(Z) = -{v\over 4 i} \left[\log {\tilde{K} - iZ\over \tilde{K} + iZ} + i \log {\tilde{K} + Z\over \tilde{K} - Z} \right]
 }
and $\alpha \equiv v K_1/\tilde{K}$. Equation \eqref{E:maindiffeq} can be solved perturbatively at large $K$ using methods developed in \cite{Yarom:2007ap,Gubser:2007nd}.\footnote{In \cite{Yarom:2007ap} the large momentum asymptotics were computed to order ${\cal O}(T^2)$ using the WKB method which required a piecewise approximation to the Schr\"odinger potential.  This approximation breaks down at order $\mathcal{O}(T^4)$ so instead we use an iterative scheme developed in \cite{Gubser:2007nd} for extracting the corresponding correction.}

Writing
$\Phi_S=\sum \tilde{K}^{-n} \psi_n$ and solving the equations of motion perturbatively for each $\psi_n$, one finds
\begin{equation}
	\psi_0 = \frac{1}{2}\pi Z^2 (C_1+C_2)(\StruveL{0}(Z)-\BesselI{0}(Z))+C_2 Z\,,
\end{equation}
where $\BesselI{0}$ is a modified Bessel function of the first kind, $\StruveL{0}$ is a modified Struve function, and $C_{1, 2}$ are given by
\begin{equation}
    C_1 =\frac{\alpha \,\left( -2 + {\alpha }^2 \right) \,
            \left( 2 - {\alpha }^2 +
            v^2\,\left( 1 + {\alpha }^2 \right)  \right) }
        {1 +{\alpha }^2}\qquad
    C_2 =\frac{3\,\alpha \,\left( 2 - {\alpha }^2 +
            v^2\,\left( 1 + {\alpha }^2 \right)  \right) }
        {1 +{\alpha }^2}\,.
\end{equation}
At the next order one has
 \eqn{psi2Def}{
	\psi_2 = C_3 Z^2+C_4 Z^4\,,
 }
with
\begin{equation}
    C_4 = -\frac{1}{3}i C_1 \alpha\qquad C_3 =
        -\frac{2}{3}i\,\left( -5\,{\alpha }^2 - 2\,{\alpha }^4 +
            v^2\,\left( 1 + {\alpha }^2 \right) \,
            \left( 9 + 2\,{\alpha }^2 \right)  \right)\,.
\end{equation}

$\psi_4$ can be found using the Green's functions method. Consider
\begin{equation}
\label{E:EOMpsi4S}
	-\psi_4 = Z^2 \BesselK{0}(Z) \int_0^Z dx \, x^{-1} \BesselI{0}(x) J(x)  + Z^2 \BesselI{0}(Z) \int_Z^{K} dx \, x^{-1} \BesselK{0}(x) J(x)\,,
\end{equation}
where now
 \eqn{JDef}{
	J(x) = -2\pi(C_1+C_2)Z^5 A_1(Z)
       +A_0(Z) \sum_i \delta_i Z^i
        +\sum_i \epsilon_i Z^i
 }
and
\begin{align}
    \delta_2 &=\frac{12\,\left( C_1 + C_2 \right) \,\pi }
        {1 + {\alpha }^2}\\
    \delta_4 &=4\,\left( C_1 + C_2 \right) \,\pi\\
    \delta_6 &=\frac{\left( C_1 + C_2 \right) \,\pi \,
        \left( -1 + {\alpha }^4 \right) }{2\,
        \left( 1 + {\alpha }^2 \right) }\\
    \epsilon_1 &=\frac{6\,\left( -4\,C_2\,
        \left( 1 + {\alpha }^2 \right)  +
        9\,\alpha \,\left( 2 - {\alpha }^2 +
        v^2\,\left( 1 + {\alpha }^2 \right)  \right)  \right) }
        {{\left( 1 + {\alpha }^2 \right) }^2}\\
    \epsilon_3 &=\frac{-6\,\left( 2\,C_2\,
       {\left( 1 + {\alpha }^2 \right) }^2 -
       \alpha \,\left( 8 - 11\,{\alpha }^2 - 10\,{\alpha }^4 +
       v^2\,\left( 7 + 11\,{\alpha }^2 + 4\,{\alpha }^4 \right)
       \right)  \right) }{{\left( 1 + {\alpha }^2 \right) }^2}\\
    \epsilon_5 &=
        \frac{\alpha \,\left( 60 - 8\,{\alpha }^2 + 7\,{\alpha }^4 +
        v^2\,\left( 18 + 11\,{\alpha }^2 - 7\,{\alpha }^4 \right)
        \right) }{6\,\left( 1 + {\alpha }^2 \right) }\\
    \epsilon_7 &=\frac{-\left( {\alpha }^3\,\left( -2 + {\alpha }^2 \right) \,
        \left( 2 - {\alpha }^2 +
        v^2\,\left( 1 + {\alpha }^2 \right)  \right)  \right) }
        {18\,\left( 1 + {\alpha }^2 \right) }\,.
\end{align}

Since for small values of their arguments $J \sim x^1$, $\BesselI{0} = 1+\mathcal{O}\left(Z^2\right)$ and $\BesselK{0} \sim 1 +\ln Z +\mathcal{O}\left(Z^2\right)$, one has
 \eqn{psi4Expanded}{
	- \left(\frac{-12 \pi \mathcal{E} \alpha}{\tilde{K}}+i\frac{6 v^2}{\tilde{K}^2}\right) =
	\left(
		-\frac{1}{2}\pi(C_1+C_2)
		+C_3
		-\frac{1}{\tilde{K}^4}\int_0^{\infty} dx \, x^{-1} \BesselK{0}(x) J(x)
		\right)+\mathcal{O}(\tilde{K}^{-5})
 }
where the upper limit of the second integral in \eqref{E:EOMpsi4S} has been taken to infinity.

Carrying out the integral, one finds
\begin{multline}
\label{E:QElargeK}
	-\pi\mathcal{E} = \frac{\pi \tilde{K}
            \left( 2 - {\alpha }^2 +
            v^2\,\left( 1 + {\alpha }^2 \right)  \right) }{24}
            +
        i\frac{\alpha
            \left( -5 - 2\,{\alpha }^2 +
            v^2\,\left( 11 + 2\,{\alpha }^2 \right)  \right) }{
            18 \tilde{K}}
            \\
            -
        \frac{\pi \,\left( 7 + 3\,{\alpha }^2 \right) \,
            \left( 2 - {\alpha }^2 +
            v^2\,\left( 1 + {\alpha }^2 \right)  \right) }{24\,
            \tilde{K}^3} + \mathcal{O}(\tilde{K}^{-5})\,.
\end{multline}

\section{Numerical Fourier transforms}
\label{FFT}

In this appendix we describe our conventions for numerical Fourier transforms and how they relate to the underlying code, namely Mathematica's implementation of the FFT\@.  Although the contents of this section are entirely elementary, we found it a useful reference.  Using the default settings, Mathematica's FFT code transforms a list $u_r$ of complex numbers with $1 \leq r \leq N$ into another list $v_s$ according to
 \eqn{MathFourier}{
  v_s = {1 \over \sqrt{N}} \sum_{r=1}^N u_r
    e^{2\pi i (r-1)(s-1)/N} \,.
 }
(We discuss here only the case of a one-dimensional FFT; higher-dimensional cases follow in the expected fashion.)  Consider now a real function
 \eqn{fXfK}{
  f(x) = \int_{-\infty}^\infty {dk \over 2\pi}
   e^{ikx} \hat{f}(k)
 }
which is peaked at $x=x_0$ and decreases to $0$ on a length scale $\Delta x$.  The Fourier transform $\hat{f}(k)$ satisfies $\hat{f}(-k) = \hat{f}(k)^*$, and for smooth $f(x)$, it decreases to $0$ roughly on a scale $\Delta k = 1/\Delta x$.  Taking this as a defining relation for $\Delta k$, let us further define
 \eqn{SeveralDefs}{
  \delta k = \sqrt{2\pi \over N} \Delta k \qquad
  \delta x = \sqrt{2\pi \over N} \Delta x \,.
 }
Using the following relations between $r,s$ and $k,x$ (with $N$ assumed to be even):
 \eqn{rskxRel}{
  k &= \left\{
   \seqalign{\span\TL &\qquad\span\TT}{
    \left( r - {1 \over 2} \right) \delta k & for
     $1 \leq r \leq N/2$  \cr
    \left( -N + r - {1 \over 2} \right) \delta k & for
     $N/2 < r \leq N$
   } \right.  \cr\noalign{\vskip2\jot}
  x-x_0 &= \left\{
   \seqalign{\span\TL &\qquad\span\TT}{
    \left( s - {1 \over 2} \right) \delta x & for
     $1 \leq s \leq N/2$  \cr
    \left( -N + s - {1 \over 2} \right) \delta x & for
     $N/2 < s \leq N$
   } \right.
 }
and the following values for $u_r$ and $v_s$:
 \eqn{vsValues}{
  u_r = \hat{f}(k) e^{ikx_0 + {i \over 2} k \delta x -
    {i \over 4} \delta k \delta x} \qquad
  v_s = \delta x \sqrt{N} f_N(x) e^{-{i \over 2}
    \delta k (x-x_0)} \,,
 }
one can show starting from \eno{MathFourier} that $f_N(x) \to f(x)$ as $N \to \infty$.  It's useful to define
 \eqn{kmaxEtc}{
  k_{\rm low} &= -k_{\rm high} = {N-1 \over 2} \delta k =
   (N-1) \sqrt{\pi \over 2N} \Delta k  \cr
  x_{\rm low} &= x_0 - {N-1 \over 2} \delta x
    = x_0 - (N-1) \sqrt{\pi \over 2N} \Delta x  \cr
  x_{\rm high} &= x_0 + {N-1 \over 2} \delta x
    = x_0 + (N-1) \sqrt{\pi \over 2N} \Delta x
 }
because then the evaluations of $\hat{f}(k)$ occur at $N$ points, spaced by $\delta k$, the first of which is $k_{\rm low}$ and the last of which is $k_{\rm high}$; and likewise $f_N(x)$ is defined at $N$ points spaced by $\delta x$, the first of which is $x_{\rm low}$ and the last of which is $x_{\rm high}$.

\clearpage
\bibliographystyle{ssg}
\bibliography{fourier}

\end{document}